\documentclass[final,12pt]{elsarticle}

\usepackage{amssymb}
\usepackage{amsmath}
\usepackage[margin=0.97in]{geometry}
\usepackage[dvipsnames]{xcolor}
\usepackage{mdframed}
\usepackage[hyphens]{url}
\usepackage[hidelinks]{hyperref}
\usepackage{algorithm}
\usepackage{algpseudocode}
\usepackage{enumitem}
\usepackage{subcaption}
\usepackage{soul}
\usepackage{flafter}
\usepackage{placeins}

\journal{}

\begin{document}

\begin{frontmatter}

\title{\texorpdfstring{A fast incompressible Navier--Stokes solver for non-uniform grids}{A fast incompressible Navier--Stokes solver for non-uniform grids}}

\author[inst1]{Pedro Costa}\corref{corr}\ead{P.SimoesCosta@tudelft.nl}
\cortext[corr]{Corresponding author.}
\author[inst1]{Duarte Palancha}
\author[inst2]{Joshua Romero}
\author[inst3]{Roberto Verzicco}
\author[inst2]{Massimiliano Fatica} %
\affiliation[inst1]{organization={Process \& Energy Department, Delft University of Technology},%
            addressline={Leeghwaterstraat 39}, 
            city={Delft},
            postcode={2628 CB},
            country={The Netherlands}}

\affiliation[inst2]{organization={NVIDIA Corporation},%
            addressline={2788 San Tomas Expressway}, 
            city={Santa Clara},
            postcode={95051}, 
            state={CA},
            country={USA}}

\affiliation[inst3]{organization={Gran Sasso Science Institute},%
            addressline={Viale Francesco Crispi, 7; Rectorate, Via Michele Iacobucci, 2}, 
            city={L'Aquila},
            postcode={67100}, 
            country={Italy}}

\begin{abstract}
We present a scalable incompressible Navier--Stokes solver for three-dimensional Cartesian grids with non-uniform spacing. A direct tensor-product--Thomas method solves the constant-coefficient Poisson and Helmholtz equations arising from pressure projection and implicit diffusion, without approximate factorization. On uniform grids, the method recovers the classical eigenfunction-expansion method evaluated with fast Fourier transforms (FFTs). On stretched grids, diagonal scaling symmetrizes the one-dimensional Laplace operators, and the resulting numerical eigenbasis transforms are evaluated as general matrix--matrix multiplications (GEMMs). FFTs and GEMMs can be selected independently in each diagonalized direction while retaining the pencil decomposition, collective transposes, and tridiagonal machinery of an established FFT-based solver. The elliptic solver is verified to round-off accuracy, and the complete flow solver is validated against benchmark flows. Against geometric multigrid and block cyclic reduction with FFT diagonalization, the present method achieves the lowest time-to-solution among the tested approaches; at fixed grid dimensions, its cost is insensitive to grid stretching. CPU and multi-GPU tests show that GEMM-rich variants attain higher strong-scaling efficiency by better amortizing communication. On a single GPU, the fully GEMM-based variant increases the Poisson cost by $2.8\times$ but the complete Navier--Stokes step cost by only $1.8\times$. Weak scaling exposes the trade-off: dense-transform costs grow with the global transform dimension, whereas non-uniform meshes can reduce the required number of grid points. The resulting open-source solver, \texttt{CaNS-EIGEN}, extends transform-based direct solution to large-scale simulations on grids stretched in multiple directions.
\end{abstract}


\end{frontmatter}

\section{Introduction}
\label{sec:introduction}

Efficient elliptic solvers lie at the heart of large-scale simulations of incompressible flow. Projection and pressure-correction methods enforce the divergence-free constraint via a pressure Poisson equation \cite{Chorin-MC-1968,Kim-and-Moin-JCP-1985,Rai-and-Moin-JCP-1991}, whose solution often constitutes a substantial portion of the computational cost. On stretched meshes, the small cells often required near walls impose a severe viscous-stability restriction when diffusion is treated explicitly. Treating diffusion implicitly removes this restriction, but introduces Helmholtz equations. Improvements in the underlying Poisson and Helmholtz solution procedures therefore translate directly into faster incompressible-flow solvers. The need is particularly acute in direct numerical simulations (DNS) of turbulence, where billions of degrees of freedom or more may be needed to resolve the relevant scales \cite{Ishihara-et-al-ARFM-2009,Pirozzoli-et-al-JFM-2021,Yeung-et-al-JFM-2025}.

Second-order finite-difference (or finite-volume) methods on staggered, structured grids are widely used for DNS of incompressible flows. Following seminal work \cite{Kim-and-Moin-JCP-1985,Verzicco-and-Orlandi-JCP-1996}, these methods have been shown to reproduce canonical turbulent flows with fidelity comparable to pseudo-spectral and higher-order methods when adequate resolution is employed \cite{Vreman-and-Kuerten-PoF-2014,Moin-and-Verzicco-EJMBF-2016,Pirozzoli-et-al-JFM-2021}. They also retain flexibility in boundary conditions and grid-point distribution and provide an efficient base for embedded-boundary techniques, including immersed-boundary methods for complex geometries \cite{Fadlun-et-al-JCP-2000,Breugem-and-Boersma-PoF-2005,Uhlmann-JCP-2005} and interface-capturing methods for multi-fluid flows \cite{Tryggvason-et-al-2011}.

The performance of these structured-grid methods at DNS scale depends critically on their elliptic solver. This requirement has become increasingly pressing as high-performance computing has shifted toward heterogeneous, GPU-dominated architectures \cite{Roccon-et-al-PRF-2026}. Although block cyclic reduction (BCR) \cite{Sweet-SIAM-1974} and geometric multigrid methods \cite{Trottenberg-et-al-2001} are highly effective on CPUs, their recursive reduction or coarsening stages do not always map naturally onto modern accelerators. These architectures benefit from exposing many thousands of concurrent threads, whereas successive reduction levels produce progressively smaller local problems and may therefore under-utilize GPU resources.

In relatively simple domains, highly efficient direct Poisson solvers based on eigenfunction expansions provide an attractive alternative. When at least two directions have uniform grid spacing and simple boundary conditions, the discrete Laplacian can be diagonalized in these directions using Fourier-based transforms, reducing the original problem to a set of decoupled one-dimensional tridiagonal systems \cite{Hockney-JACM-1965,Buzbee-et-al-SIAM-1970,Swarztrauber-SIAM-1977,Schumann-and-Sweet-JCP-1988}. This approach underpins several established DNS solvers, such as \texttt{AFiD} \cite{Van-der-Poel-et-al-CF-2015} and \texttt{CaNS} \cite{Costa-CAMWA-2018}, which have demonstrated strong performance and scalability on both CPU- and GPU-based architectures \cite{Costa-et-al-CAMWA-2021,Zhu-et-al-CPC-2018}. In these solvers, batches of one-dimensional discrete Fourier/sine/cosine transforms are computed efficiently using FFT libraries, enabling the fast solution of the resulting tridiagonal systems. Despite the collective (\emph{all-to-all}) communications associated with the pencil transposes required by these solvers, modern communication libraries such as \texttt{2DECOMP\&FFT} \cite{Li-and-Laizet-CRAY-2010,Rolfo-et-al-JOSS-2023} and \texttt{cuDecomp} \cite{Romero-et-al-PASC-2022} allow them to tackle problems on modern supercomputers at ever-increasing scales.

However, a key limitation of FFT-based direct finite-difference Poisson solvers is their reliance on uniform grids in the directions where the transforms are applied. To accommodate non-uniform grids, one often resorts to other approaches such as geometric multigrid methods, or classical direct solvers that combine FFT-based diagonalization along one direction with BCR \cite{FISHPACK,Swarztrauber-and-Sweet-JCAM-1989} (see also \cite{Borrell-et-al-JCP-2011,Costa-CPC-2022} for other applications of FFT-based diagonalization). On strongly non-uniform grids, however, geometric multigrid may become significantly more expensive unless its smoothing and coarsening strategies are adapted to the induced anisotropy \cite{Trottenberg-et-al-2001}. Without such adaptation, more iterations or cycles may be needed to reach a given tolerance. In the context of an incompressible Navier--Stokes solver, this iterative error behavior may lead to poorer enforcement of the divergence-free condition as the grid stretching becomes more severe. Moreover, as noted above, these alternatives are not ideally suited for current GPU architectures.

The limitations of these solvers are particularly visible in wall-turbulence DNS with more than one inhomogeneous direction. A recent survey of GPU-accelerated turbulence simulations reports maximum friction Reynolds numbers of approximately $10\,000$--$12\,000$ for channels and pipes, but only about $2\,000$ for square ducts; boundary-layer and duct calculations likewise remain well below their channel and pipe counterparts \cite{Roccon-et-al-PRF-2026}. Several geometric and numerical difficulties contribute to this gap, but the loss of a second homogeneous direction also removes the efficient double-transform Poisson inversion available in channels and pipes. Modesti and Pirozzoli \cite{Modesti-and-Pirozzoli-JFM-2022}, for example, used a compressible solver at low Mach number for a physically incompressible square-duct flow and explicitly identified this loss as a numerical obstacle. Direct incompressible calculations have instead combined an FFT along the homogeneous direction with a two-dimensional \texttt{BLKTRI} solve in the cross-section \cite{Orlandi-and-Pirozzoli-JoT-2020}; the remaining two-dimensional inversion, however, does not retain the scalability of the double-transform case. These limitations motivate fast, \emph{direct} Poisson and Helmholtz solvers for incompressible flow that accommodate non-uniform grids and expose compute-intensive kernels well suited to GPUs.

Lynch et al.~\cite{Lynch-et-al-1964} described a tensor-product method for the direct solution of separable difference equations. On uniform grids, this construction reduces to the familiar Fourier eigenfunction expansions used in fast direct elliptic solvers \cite{Swarztrauber-SIAM-1977}. Farnell \cite{Farnell-JCP-1980} extended it to non-uniform grids by generating the eigenvectors numerically in each stretched direction. The same matrix-decomposition idea was subsequently developed for broader classes of separable boundary-value problems \cite{Bialecki-and-Fairweather-JCPAM-1993,Bialecki-et-al-NA-2011}. Vitoshkin and Gelfgat~\cite{Vitoshkin-and-Gelfgat-CCP-2013}, for instance, used tensor-product factorization to invert Laplace and Helmholtz operators within a Stokes-operator preconditioner for Newton and Arnoldi calculations. More recently, Krasnov et al.~\cite{Krasnov-et-al-JCP-2023} applied a closely related tensor-product--Thomas approach in a liquid-metal magnetohydrodynamics solver, evaluating the numerical eigenbasis transforms as matrix--matrix multiplications.

Building on Farnell's matrix decomposition \cite{Farnell-JCP-1980}, we use a common direct tensor-product framework for pressure projection and selectable implicit diffusion in one, two, or all three directions. The multidirectional Helmholtz problems are solved without approximate factorization. We embed this framework in the pencil-decomposed, GPU-accelerated \texttt{CaNS} solver \cite{Costa-CAMWA-2018,Costa-et-al-CAMWA-2021}. FFT- and GEMM-based transforms can be selected independently in each diagonalized direction, so the same implementation accommodates large-scale simulations on grids stretched in multiple directions. We then compare the method with geometric multigrid and block cyclic reduction, and examine its scaling on modern CPU and multi-GPU systems, including a cluster equipped with a high-bandwidth multinode GPU fabric. The resulting solver, \texttt{CaNS-EIGEN}, is freely available and open-source at \url{https://github.com/CaNS-World/CaNS-EIGEN}.

We validate the complete solver against established laminar and turbulent benchmarks, including cases with explicit and implicit diffusion. In the performance tests, the present method achieves the lowest Poisson-solve time among the geometric multigrid and block-cyclic-reduction solvers considered here and, at fixed grid dimensions, remains insensitive to mesh grading. The multi-GPU results show that the GEMM-based variants better amortize communication than the FFT-based variants. Tests with and without Multinode NVLink (MNNVL) also show how the inter-node network affects scaling. Finally, the weak-scaling results expose the main trade-off of the method: dense transforms become more expensive as the global problem grows, but mesh stretching can reduce the number of grid points required.

The remainder of this paper is organized as follows. Section~\ref{sec:numerical-method} introduces the Navier--Stokes discretization and derives the common Poisson/Helmholtz solution framework. Section~\ref{sec:implementation} describes the CPU/GPU implementation and parallelization strategy. Section~\ref{sec:results} presents verification, flow validation, and performance results. Finally, Section~\ref{sec:conclusions-and-outlook} summarizes the main findings and outlines directions for future work.

\section{Numerical method}
\label{sec:numerical-method}

\subsection{Navier--Stokes solver}\label{sec:Navier--Stokes-solver}

We solve the incompressible Navier--Stokes equations
\begin{subequations}\label{eq:ns}
\begin{align}
  \boldsymbol{\nabla} \cdot \mathbf{u} &= 0,\\
  \partial_t \mathbf{u} + \boldsymbol{\nabla}\cdot(\mathbf{u} \otimes \mathbf{u})
  &= -\boldsymbol{\nabla} P + \nu \nabla^2 \mathbf{u},
\end{align}
\end{subequations}
where $\mathbf{u}$ is the velocity vector, $P$ is the kinematic pressure, and $\nu$ is the kinematic viscosity. Here, $\otimes$ denotes the outer product of vectors; the same symbol is used for the Kronecker product of matrices in Section~\ref{sec:elliptic-solver}.

The spatial discretization uses second-order finite differences on non-uniform, structured, staggered Cartesian grids \cite{Harlow-and-Welch-PoF-1965}. The staggered arrangement avoids odd--even decoupling between pressure and velocity, while the spatial discretization preserves kinetic energy in a semi-discrete sense \cite{Verstappen-and-Veldman-JCP-2003}. Time advancement uses Wray's low-storage three-stage Runge--Kutta (RK3) scheme within a fractional-step method \cite{Chorin-MC-1968,Wray-1990,Kim-and-Moin-JCP-1985}.

To allow diffusion to be treated explicitly or implicitly by direction, we decompose the discrete Laplacian as $\mathcal{L}=\mathcal{L}_E+\mathcal{L}_I$, where the subscripts $E$ and $I$ denote the explicitly and implicitly treated contributions, respectively, and $\mathcal{L}_\alpha$ denotes the contribution along direction $\alpha\in\{x,y,z\}$. We consider four choices for $\mathcal{L}_I$: $\mathcal{L}_I=0$ (fully explicit), $\mathcal{L}_I=\mathcal{L}_z$ (one implicit direction, $z$), $\mathcal{L}_I=\mathcal{L}_y+\mathcal{L}_z$ (two implicit directions, $yz$), and $\mathcal{L}_I=\mathcal{L}_x+\mathcal{L}_y+\mathcal{L}_z=\mathcal{L}$ (all three implicit directions, $xyz$). When $z$ is the wall-normal direction, the $z$-only choice corresponds to the established practice of treating diffusion implicitly only in the refined wall-normal direction, while avoiding the communication required to include the wall-parallel directions in the implicit solve \cite{Van-der-Poel-et-al-CF-2015}.

Each time step is advanced through three Runge--Kutta substeps, indexed by $k\in\{1,2,3\}$. The coefficients are $(\alpha_1,\alpha_2,\alpha_3)=(8/15,5/12,3/4)$ and $(\beta_1,\beta_2,\beta_3)=(0,-17/60,-5/12)$, with $\gamma_k=\alpha_k+\beta_k$. The stage velocities are defined such that $\mathbf{u}^{1}=\mathbf{u}^{n}$ at the beginning of the time step and $\mathbf{u}^{4}=\mathbf{u}^{n+1}$ after the third substep:
\begin{subequations}\label{eq:rk3}
\begin{align}
    \mathbf{u}^{**}
    &= \mathbf{u}^k + \Delta t\left[\alpha_k(\mathcal{A}\mathbf{u}^{k}+\nu\mathcal{L}_E\mathbf{u}^{k}) + \beta_k(\mathcal{A}\mathbf{u}^{k-1} + \nu\mathcal{L}_E\mathbf{u}^{k-1})\right.\notag\\
    &\left.\qquad\qquad\quad + \gamma_k(-\mathcal{G} P^{k-1/2}+\nu\mathcal{L}_I\mathbf{u}^k)\right],\label{eqn:predictor_ns}\\
    \left(\mathcal{I}-\frac{\gamma_k\nu\Delta t}{2}\mathcal{L}_I\right)\mathbf{u}^*
    &= \mathbf{u}^{**}-\frac{\gamma_k\nu\Delta t}{2}\mathcal{L}_I\mathbf{u}^k,\label{eqn:helmholtz_ns}\\
    \mathcal{L}\Phi
    &= \frac{\mathcal{D}\mathbf{u}^*}{\gamma_k\Delta t},\label{eqn:poi_ns}\\
    \mathbf{u}^{k+1}
    &= \mathbf{u}^* - \gamma_k\Delta t \mathcal{G}\Phi,\\
    P^{k+1/2}
    &= P^{k-1/2} + \Phi - \frac{\gamma_k\nu\Delta t}{2}\mathcal{L}_I \Phi.
\end{align}
\end{subequations}
The quantities $P^{k-1/2}$ and $P^{k+1/2}$ denote the pressure before and after substep $k$, respectively.
Here, $\mathcal{I}$ is the identity operator, and $\mathcal{A}\mathbf{u}$ denotes the discrete form of the negative advection term, $-\boldsymbol{\nabla}\cdot(\mathbf{u}\otimes\mathbf{u})$. The operators $\mathcal{L}$, $\mathcal{G}$, and $\mathcal{D}$ denote the discrete Laplacian, gradient, and divergence, respectively. The velocity $\mathbf{u}^{**}$ is the intermediate predictor, $\mathbf{u}^{*}$ is the velocity after the implicit diffusion solve, and $\Phi$ is the pressure correction.

For $\mathcal{L}_I=0$, Eq.~\eqref{eqn:helmholtz_ns} reduces to $\mathbf{u}^{*}=\mathbf{u}^{**}$, and Eq.~\eqref{eqn:predictor_ns} recovers the fully explicit momentum update. For $\mathcal{L}_I\neq0$, the selected diffusion terms are treated with the Crank--Nicolson method, and Eq.~\eqref{eqn:helmholtz_ns} defines a Helmholtz problem for each velocity component. For the two- and three-direction choices, we solve this multidimensional problem without approximate factorization. Approximate factorization would replace it with a sequence of one-dimensional solves \cite{Peaceman-and-Rachford-1955,Kim-and-Moin-JCP-1985}, but would introduce cross-directional splitting terms of order $(\nu\Delta t)^2$ and, in three dimensions, an additional term of order $(\nu\Delta t)^3$ \cite{Douglas-NM-1962,Beam-and-Warming-JCP-1976}. These terms can become appreciable when diffusion is stiff in more than one direction, as in low-Reynolds-number flows, strongly stretched meshes, or calculations using large implicit time steps. The unsplit solve avoids them.

\subsection{Direct Poisson/Helmholtz solver}\label{sec:elliptic-solver}

\subsubsection{One-dimensional operator and eigendecomposition}

\begin{figure}
    \centering
    \includegraphics[width=0.55\linewidth]{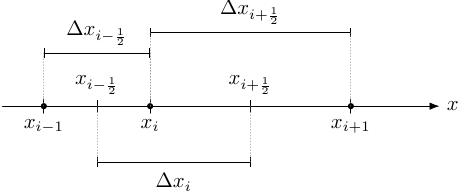}
    \caption{Non-uniform one-dimensional grid and spacing definitions. Black circles denote the nodal locations $x_i$ associated with the unknowns $\phi_i$, while $x_{i\pm 1/2}$ indicate the half-index locations.}
    \label{fig:nonuniform_grid_sketch}
\end{figure}

We begin with the one-dimensional Poisson equation, $\partial_{xx}\phi=f$, discretized with second-order central finite differences at the $N=N_x$ grid points $x_i$, $i=1,\ldots,N$, shown in Fig.~\ref{fig:nonuniform_grid_sketch}. The resulting system is tridiagonal and could be solved directly with the Thomas algorithm. Here, however, its eigenvectors provide the basis for diagonalizing selected directions of the multidimensional Poisson and Helmholtz operators, leaving independent tridiagonal systems in the remaining direction.
At an interior grid point $x_i$, the discretization is
\begin{equation}
  \frac{1}{\Delta x_i}
  \left(
    \frac{\phi_{i+1}-\phi_i}{\Delta x_{i+1/2}}
    -
    \frac{\phi_i-\phi_{i-1}}{\Delta x_{i-1/2}}
  \right)
  = f_i,
  \label{eqn:poisson-fd2}
\end{equation}
where $\Delta x_{i+1/2}=x_{i+1}-x_i$, $x_{i+1/2}=(x_{i+1}+x_i)/2$, and $\Delta x_i=x_{i+1/2}-x_{i-1/2}$. This yields the linear system $\boldsymbol{T}\boldsymbol{\phi}=\boldsymbol{f}$, where $\boldsymbol{T}\in\mathbb{R}^{N\times N}$ and $\boldsymbol{\phi},\boldsymbol{f}\in\mathbb{R}^{N}$. For non-periodic boundary conditions, $\boldsymbol{T}$ has the tridiagonal form
\begin{equation}
  \boldsymbol{T} =
  \begin{pmatrix}
    b_1 & c_1 & 0 & \cdots & 0 \\
    a_2 & b_2 & c_2 & \cdots & 0 \\
    \vdots & \vdots & \vdots & \ddots & \vdots \\
    0 & \cdots & a_{N-1} & b_{N-1} & c_{N-1}\\
    0 & \cdots & 0 & a_N & b_N
  \end{pmatrix},
  \qquad
  \begin{aligned}
    a_i &= \left(\Delta x_i\,\Delta x_{i-1/2}\right)^{-1},\\
    c_i &= \left(\Delta x_i\,\Delta x_{i+1/2}\right)^{-1},\\
    b_i &= -\left(a_i+c_i\right).
  \end{aligned}
  \label{eqn:tridiagonal-coefficients}
\end{equation}
The coefficients of the first and last rows are modified as required by the boundary conditions; periodic boundaries also introduce cyclic corner entries.

The tensor-product construction uses the eigendecomposition
$\boldsymbol{T}=\boldsymbol{Q}\boldsymbol{\Lambda}\boldsymbol{Q}^{-1}$
\cite{Lynch-et-al-1964,Haidvogel-et-al-JCP-1979,Bialecki-and-Fairweather-JCPAM-1993}, where $\boldsymbol{\Lambda}$ is the diagonal matrix of eigenvalues and $\boldsymbol{Q}$ contains the corresponding eigenvectors. On a non-uniform grid, $\boldsymbol{T}$ is generally non-symmetric, and its eigenvectors are generally non-orthogonal, so that $\boldsymbol{Q}^{-1}\neq\boldsymbol{Q}^{T}$. This prevents the direct use of efficient symmetric-tridiagonal eigensolvers.

The operator nevertheless has a weighted symmetry. Define the cell-size matrix
$\boldsymbol{D}=\operatorname{diag}(\Delta x_1,\Delta x_2,\ldots,\Delta x_N)$.
Multiplying Eq.~\eqref{eqn:poisson-fd2} by $\Delta x_i$ yields the row-scaled matrix $\boldsymbol{D}\boldsymbol{T}$, which is symmetric because
$\Delta x_i c_i=\Delta x_{i+1}a_{i+1}$.
The row-scaled system is also the finite-volume form of the discrete Laplace equation. It follows that $\boldsymbol{T}$ is similar to the symmetric matrix \cite{Farnell-JCP-1980}
\begin{equation}
  \boldsymbol{\tilde{T}}
  = \boldsymbol{D}^{1/2}\boldsymbol{T}\boldsymbol{D}^{-1/2}.
  \label{eqn:reverse-similarity-transform}
\end{equation}

Because Eq.~\eqref{eqn:reverse-similarity-transform} is a similarity transformation, $\boldsymbol{\tilde{T}}$ has the same eigenvalues as $\boldsymbol{T}$. Since $\boldsymbol{\tilde{T}}$ is symmetric, it admits the eigendecomposition
$\boldsymbol{\tilde{T}}=\boldsymbol{\tilde{Q}}\boldsymbol{\Lambda}\boldsymbol{\tilde{Q}}^{T}$,
where $\boldsymbol{\tilde{Q}}$ is orthogonal. The eigenvector matrices of the original operator are then
\begin{equation}
  \boldsymbol{Q}
  = \boldsymbol{D}^{-1/2}\boldsymbol{\tilde{Q}},
  \qquad
  \boldsymbol{Q}^{-1}
  = \boldsymbol{\tilde{Q}}^{T}\boldsymbol{D}^{1/2}.
  \label{eqn:q-from-tildeq}
\end{equation}

More generally, any tridiagonal matrix satisfying $c_i a_{i+1}>0$ can be symmetrized by a diagonal similarity transformation \cite{Kreer-SPA-1994}; the cell-size scaling above is the form obtained for the present discrete Laplace operator.

For a fixed grid and set of boundary conditions, the eigendecomposition is computed once and reused for every right-hand side. The one-dimensional procedure is summarized as follows:
\begin{mdframed}[backgroundcolor=gray!10,linewidth=0.3pt,nobreak=true]
\small
\textbf{1.~Initialization}
\begin{enumerate}[label=1.\arabic*]\itemsep0em
  \item Form the symmetric operator: $\boldsymbol{\tilde{T}}\leftarrow\boldsymbol{D}^{1/2}\boldsymbol{T}\boldsymbol{D}^{-1/2}$.
  \item Compute its eigendecomposition: $\boldsymbol{\tilde{T}}=\boldsymbol{\tilde{Q}}\boldsymbol{\Lambda}\boldsymbol{\tilde{Q}}^{T}$.
  \item Form the forward and inverse transforms: $\boldsymbol{Q}\leftarrow\boldsymbol{D}^{-1/2}\boldsymbol{\tilde{Q}}$ and $\boldsymbol{Q}^{-1}\leftarrow\boldsymbol{\tilde{Q}}^{T}\boldsymbol{D}^{1/2}$.
\end{enumerate}
\textbf{2.~Solution}
\begin{enumerate}[label=2.\arabic*]\itemsep0em
  \item Transform the right-hand side: $\boldsymbol{\hat{f}}\leftarrow\boldsymbol{Q}^{-1}\boldsymbol{f}$.
  \item Solve in the eigenbasis: $\boldsymbol{\hat{\phi}}\leftarrow\boldsymbol{\Lambda}^{-1}\boldsymbol{\hat{f}}$.
  \item Transform back to physical space: $\boldsymbol{\phi}\leftarrow\boldsymbol{Q}\boldsymbol{\hat{\phi}}$.
\end{enumerate}
\end{mdframed}
The diagonal inverse above assumes a non-singular operator; null modes associated with periodic or Neumann--Neumann boundary conditions are treated separately in the multidimensional formulation below.

Finally, on a uniform grid, $\boldsymbol{T}$ is symmetric, and the Fourier-based solution \cite{Swarztrauber-SIAM-1977} is a special case of the procedure above. The eigenvalues (or modified wavenumbers) can then be obtained analytically for several combinations of boundary conditions \cite{Swarztrauber-and-Sweet-JCAM-1989,Schumann-and-Sweet-JCP-1988}, while the projections $\boldsymbol{Q}^{-1}\boldsymbol{f}$ and $\boldsymbol{Q}\boldsymbol{\hat{\phi}}$ become discrete trigonometric transforms that can be evaluated with FFTs.

\subsubsection{Three-dimensional tensor-product--Thomas formulation}\label{sec:poisson-3d-solver}

We now extend the one-dimensional construction to three dimensions. Let $\boldsymbol{T}_\alpha\in\mathbb{R}^{N_\alpha\times N_\alpha}$ denote the one-dimensional discrete Poisson operator along direction $\alpha\in\{x,y,z\}$, with coefficients given by Eq.~\eqref{eqn:tridiagonal-coefficients}. Using a tensor-product ordering whose Kronecker factors correspond to $(x,y,z)$ from left to right, the three-dimensional operator acting on the vectorized field $\boldsymbol{\phi}\in\mathbb{R}^{N_xN_yN_z}$ is
\begin{equation}
  \boldsymbol{T}
  = \boldsymbol{T}_x\otimes\boldsymbol{I}_y\otimes\boldsymbol{I}_z
  + \boldsymbol{I}_x\otimes\boldsymbol{T}_y\otimes\boldsymbol{I}_z
  + \boldsymbol{I}_x\otimes\boldsymbol{I}_y\otimes\boldsymbol{T}_z,
\end{equation}
where $\otimes$ denotes the Kronecker product and $\boldsymbol{I}_\alpha$ is the identity matrix in direction $\alpha$. The discrete three-dimensional system is therefore $\boldsymbol{T}\boldsymbol{\phi}=\boldsymbol{f}$.

We diagonalize the operators along $x$ and $y$, leaving independent tridiagonal systems along $z$ \cite{Lynch-et-al-1964,Vitoshkin-and-Gelfgat-CCP-2013}. The operators $\boldsymbol{T}_x$ and $\boldsymbol{T}_y$ are symmetrized using the diagonal scalings $\boldsymbol{D}_x$ and $\boldsymbol{D}_y$, while $\boldsymbol{T}_z$ is retained in its original tridiagonal form. With $\boldsymbol{T}_\alpha=\boldsymbol{Q}_\alpha\boldsymbol{\Lambda}_\alpha\boldsymbol{Q}_\alpha^{-1}$ for $\alpha\in\{x,y\}$, the corresponding three-dimensional transformation matrices are
\begin{equation}
  \boldsymbol{Q}
  = \boldsymbol{Q}_x\otimes\boldsymbol{Q}_y\otimes\boldsymbol{I}_z,
  \qquad
  \boldsymbol{Q}^{-1}
  = \boldsymbol{Q}_x^{-1}\otimes\boldsymbol{Q}_y^{-1}\otimes\boldsymbol{I}_z.
\end{equation}
For each mode pair $(i,j)$, the transformed $x$ and $y$ operators contribute the scalar shift
\begin{equation}
  \lambda_{ij}=\lambda_{x,i}+\lambda_{y,j}.
\end{equation}

Following standard eigendecomposition-based Poisson solvers \cite{Lynch-et-al-1964,Farnell-JCP-1980,Golub-et-al-SIAM-1998}, the forward transform $\boldsymbol{Q}^{-1}\boldsymbol{f}$ is applied successively along $x$ and $y$:
\begin{subequations}\label{eq:forward-xy}
\begin{align}
  \boldsymbol{\hat{f}}^{(x)}
  &= \left(
       \boldsymbol{Q}_x^{-1}
       \otimes\boldsymbol{I}_y
       \otimes\boldsymbol{I}_z
     \right)\boldsymbol{f},\\
  \boldsymbol{\hat{f}}^{(xy)}
  &= \left(
       \boldsymbol{I}_x
       \otimes\boldsymbol{Q}_y^{-1}
       \otimes\boldsymbol{I}_z
     \right)\boldsymbol{\hat{f}}^{(x)}.
\end{align}
\end{subequations}

These transforms diagonalize the operator along $x$ and $y$, leaving $N_xN_y$ independent tridiagonal systems along $z$:
\begin{equation}
  \left(
    \lambda_{ij}\boldsymbol{I}_z+\boldsymbol{T}_z
  \right)
  \boldsymbol{\hat{\phi}}^{(xy)}_{ij}
  =
  \boldsymbol{\hat{f}}^{(xy)}_{ij},
  \qquad
  i=1,\ldots,N_x,\quad
  j=1,\ldots,N_y.
\end{equation}
Each system can then be solved directly, for example with the Thomas algorithm used in the present implementation.

The physical-space solution is recovered by applying the inverse transforms along $y$ and $x$:
\begin{subequations}\label{eq:inverse-xy}
\begin{align}
  \boldsymbol{\hat{\phi}}^{(x)}
  &= \left(
       \boldsymbol{I}_x
       \otimes\boldsymbol{Q}_y
       \otimes\boldsymbol{I}_z
     \right)\boldsymbol{\hat{\phi}}^{(xy)},\\
  \boldsymbol{\phi}
  &= \left(
       \boldsymbol{Q}_x
       \otimes\boldsymbol{I}_y
       \otimes\boldsymbol{I}_z
     \right)\boldsymbol{\hat{\phi}}^{(x)}.
\end{align}
\end{subequations}

The complete three-dimensional tensor-product--Thomas procedure is:
\begin{mdframed}[backgroundcolor=gray!10,linewidth=0.3pt,nobreak=true]
\small
\textbf{1.~Initialization}
\begin{enumerate}[label=1.\arabic*]\itemsep0em
  \item For each $\alpha\in\{x,y\}$, form the symmetric operator: $\boldsymbol{\tilde{T}}_\alpha \leftarrow \boldsymbol{D}_\alpha^{1/2}\boldsymbol{T}_\alpha\boldsymbol{D}_\alpha^{-1/2}$.
  \item Compute the eigendecomposition: $\boldsymbol{\tilde{T}}_\alpha = \boldsymbol{\tilde{Q}}_\alpha\boldsymbol{\Lambda}_\alpha\boldsymbol{\tilde{Q}}_\alpha^{T}$.
  \item Form the transforms: $\boldsymbol{Q}_\alpha\leftarrow\boldsymbol{D}_\alpha^{-1/2}\boldsymbol{\tilde{Q}}_\alpha$,\quad $\boldsymbol{Q}_\alpha^{-1}\leftarrow\boldsymbol{\tilde{Q}}_\alpha^{T}\boldsymbol{D}_\alpha^{1/2}$.
\end{enumerate}
\textbf{2.~Solution}
\begin{enumerate}[label=2.\arabic*]\itemsep0em
  \item Apply the forward transform along $x$: $\boldsymbol{\hat{f}}^{(x)} \leftarrow \left(\boldsymbol{Q}_x^{-1}\otimes\boldsymbol{I}_y\otimes\boldsymbol{I}_z\right)\boldsymbol{f}$.
  \item Apply the forward transform along $y$: $\boldsymbol{\hat{f}}^{(xy)} \leftarrow \left(\boldsymbol{I}_x\otimes\boldsymbol{Q}_y^{-1}\otimes\boldsymbol{I}_z\right)\boldsymbol{\hat{f}}^{(x)}$.
  \item Solve the tridiagonal system along $z$ for each $(i,j)$: $\left(\lambda_{ij}\boldsymbol{I}_z+\boldsymbol{T}_z\right)\boldsymbol{\hat{\phi}}^{(xy)}_{ij}=\boldsymbol{\hat{f}}^{(xy)}_{ij}$.
  \item Apply the inverse transform along $y$: $\boldsymbol{\hat{\phi}}^{(x)} \leftarrow \left(\boldsymbol{I}_x\otimes\boldsymbol{Q}_y\otimes\boldsymbol{I}_z\right)\boldsymbol{\hat{\phi}}^{(xy)}$.
  \item Apply the inverse transform along $x$: $\boldsymbol{\phi} \leftarrow \left(\boldsymbol{Q}_x\otimes\boldsymbol{I}_y\otimes\boldsymbol{I}_z\right)\boldsymbol{\hat{\phi}}^{(x)}$.
\end{enumerate}
\end{mdframed}
The computational distinction between uniform and non-uniform diagonalized directions lies in the eigenbasis transforms. For a non-uniform grid, the matrices $\boldsymbol{Q}_\alpha$ and $\boldsymbol{Q}_\alpha^{-1}$ are full, so the corresponding transforms are \emph{dense}. A forward transform along $x$, for example, comprises $N_yN_z$ matrix--vector products of size $N_x$, one for each $(j,k)$ line. By stacking these lines as columns of a matrix with leading dimension $N_x$, all products can be evaluated in a single matrix--matrix multiplication (GEMM) \cite{Krasnov-et-al-JCP-2023}. For a uniform grid in a diagonalized direction, the numerical eigenbasis transform can instead be replaced by the corresponding Fourier-based transform \cite{Kim-and-Moin-JCP-1985}. Poisson and Helmholtz systems use the same eigenvectors and transforms; the Helmholtz operator introduces only a constant shift in the modal system.

For a Poisson operator with a constant null mode, the solution is defined only up to an additive constant; the gauge condition used here is described in Section~\ref{sec:implementation}.

\section{Implementation}
\label{sec:implementation}

We implemented the method in \texttt{CaNS-EIGEN}, an extension of the \texttt{CaNS} solver \cite{Costa-CAMWA-2018,Costa-et-al-CAMWA-2021} that supports non-uniform grids in all three directions and includes the Poisson/Helmholtz framework described above. The code is openly available at \url{https://github.com/CaNS-World/CaNS-EIGEN}. It is written in modern Fortran, uses the Message-Passing Interface (\texttt{MPI}) for distributed-memory parallelization and \texttt{OpenACC} for GPU offloading, and provides CUDA Fortran interfaces to selected NVIDIA libraries. Single or double precision is selected at compile time through \texttt{C} preprocessor macros.

As in other pencil-decomposed solvers, the computational domain is partitioned along two coordinate directions. Each pencil contains complete grid lines along its orientation, allowing the corresponding directional operator to be evaluated locally. The flow solver may use any of the three pencil orientations; data are transposed only when a step of the Poisson/Helmholtz solution requires a different set of grid lines to be local to each MPI task.

\begin{figure}
  \centering
  \includegraphics[width=0.95\linewidth]{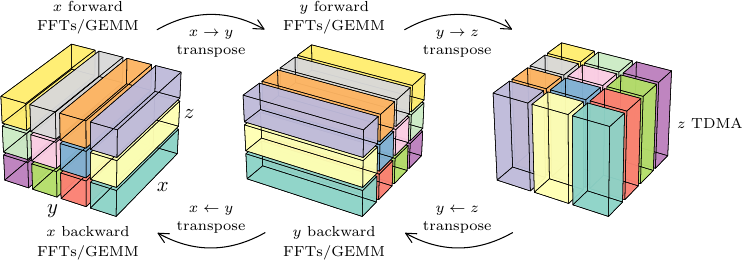}
  \caption{Pencil layouts and data redistributions in the three-dimensional Poisson/Helmholtz solver. The physical domain and coordinate axes remain fixed, while the colors indicate the distribution of data among MPI tasks. The solver follows the sequence $x\rightarrow y\rightarrow z$ for the forward transforms and tridiagonal solve, and then returns through $z\rightarrow y\rightarrow x$ for the inverse transforms. Each transpose makes the grid lines required by the next directional operation local to an MPI task. Here, TDMA denotes the Thomas algorithm for tridiagonal systems. The $z$-direction systems are solved locally in the present work, but may alternatively use a distributed TDMA algorithm \cite{Diez-et-al-CPC-2025}.}
  \label{fig:domain_decomposition}
\end{figure}

For the presentation below, we use $x$-aligned pencils as the input/output layout of the elliptic solver. Let $P_1\times P_2$ denote the two-dimensional MPI process grid, with $P_1P_2$ MPI tasks in total. Assuming an evenly distributed decomposition, let $n_\beta^\alpha$ denote the local extent in direction $\beta$ for $\alpha$-aligned pencils. The three local layouts are
\[
\begin{aligned}
x\text{-pencils}:&\quad
N_x\times n_y^x\times n_z^x,
& (n_y^x,n_z^x)
&=\left(N_y/P_1,N_z/P_2\right),\\
y\text{-pencils}:&\quad
n_x^y\times N_y\times n_z^y,
& (n_x^y,n_z^y)
&=\left(N_x/P_1,N_z/P_2\right),\\
z\text{-pencils}:&\quad
n_x^z\times n_y^z\times N_z,
& (n_x^z,n_y^z)
&=\left(N_x/P_1,N_y/P_2\right).
\end{aligned}
\]
These layouts are illustrated in Fig.~\ref{fig:domain_decomposition}.

The required collective pencil transposes are implemented with the \texttt{2DECOMP\&FFT} library on CPUs \cite{Li-and-Laizet-CRAY-2010,Rolfo-et-al-JOSS-2023} and with \texttt{cuDecomp} on NVIDIA GPUs \cite{Romero-et-al-PASC-2022}. In addition to transpose operations and halo exchanges, \texttt{cuDecomp} supports runtime autotuning of the two-dimensional process grid and communication backend. The implementation also supports \texttt{diezDecomp} \cite{Diez-et-al-CPC-2025} as an alternative backend, although it is not used in the benchmarks reported here.

The transform type is selected independently in each diagonalized direction. A uniform direction may use either FFT-based transforms or numerical eigenbasis transforms evaluated with GEMM, whereas a non-uniform direction uses the latter. The same solver pipeline supports the three implicit-diffusion choices shown in Fig.~\ref{fig:domain_decomposition}: the $xyz$ option uses the $x$ and $y$ transforms together with the $z$-direction tridiagonal solve, the $yz$ option omits the $x$ transforms, and the $z$ option performs only the tridiagonal solve.

Initialization is performed once for each directional operator. For a uniform direction, the analytical eigenvalues are stored and batched FFT plans are created. For a non-uniform direction, the symmetric transformed operator $\boldsymbol{\tilde{T}}_\alpha$ is eigendecomposed, and the resulting eigenvalues and forward/inverse transform matrices are reused for every right-hand side. Non-cyclic operators are symmetric and tridiagonal, and are decomposed using the divide-and-conquer algorithm \cite{Cuppen-NM-1980} provided by the \texttt{LAPACK} \texttt{xSTEDC} routines. Periodic couplings produce a symmetric cyclic operator, which is stored as a dense matrix and decomposed using \texttt{xSYEVD}.

This one-time initialization typically takes only a few seconds for $N_x$ or $N_y\approx5000$ on a modern CPU. At fixed $N_\alpha$, its operation count is independent of grid stretching because the direct eigensolvers do not involve a convergence iteration. For GPU runs, the eigenvectors and scaling factors are computed on the CPU and copied to the device.

During each solve, the forward and inverse numerical eigenbasis transforms are evaluated with BLAS GEMM routines, using \texttt{OpenBLAS} on CPUs and \texttt{cuBLAS} on GPUs. In a uniform direction, these operations may instead be performed with batched Fourier-based transforms using \texttt{FFTW} on CPUs and \texttt{cuFFT} on GPUs. Subsequently, for the $z$-direction systems, the benchmarks use the transpose-based local Thomas solver inherited from \texttt{CaNS} \cite{Costa-CAMWA-2018,Costa-et-al-CAMWA-2021}. The implementation also supports distributed TDMA variants \cite{Diez-et-al-CPC-2025}. In the local CPU and GPU Thomas solvers used here, the pressure gauge is imposed by detecting a numerically vanishing final pivot with a scale-aware tolerance and setting the corresponding modal unknown to zero. Algorithm~\ref{alg:poisson-solver} summarizes the complete three-dimensional workload and follows the sequence shown in Fig.~\ref{fig:domain_decomposition}.
\begin{algorithm}
\caption{Distributed-memory workload of the three-dimensional Poisson/Helmholtz solver; see Fig.~\ref{fig:domain_decomposition} and the text for the local pencil sizes. Here, ``U~Grid'' denotes a uniform grid direction, ``NU~Grid'' a non-uniform grid direction, and ``Ops.'' the computational complexity. Steps~4--6 may alternatively be replaced by a distributed TDMA algorithm \cite{Diez-et-al-CPC-2025}.}\label{alg:poisson-solver}
\begin{mdframed}[backgroundcolor=gray!10,linewidth=0.3pt]
\setlength{\fboxsep}{3pt}
\newcommand{\OneAlgoBox}[1]{%
  \noindent
  \fbox{\parbox[t]{\dimexpr\linewidth-2\fboxsep-2\fboxrule\relax}{#1}}%
}
\newcommand{\AlgoBoxTitle}[1]{%
  {\centering\textbf{#1}\par}\noindent\ignorespaces
}
\newcommand{\AlgoBoxHead}[2]{%
  {\centering\textbf{#1}~#2\par\smallskip}\noindent\ignorespaces
}
\newcommand{\AlgoSection}[1]{%
  \par\noindent\centering{\small\textbf{#1}}\par\vspace{0.01\baselineskip}%
  \noindent\footnotesize
}
\newcommand{\AlgoUniNonUni}[2]{%
  \begin{minipage}[t]{0.485\linewidth}\raggedright\textbf{NU Grid:} #2\end{minipage}\hfill
  \begin{minipage}[t]{0.485\linewidth}\raggedright\textbf{U Grid:} #1\end{minipage}%
}
\newcommand{\AlgoUniNonUniOps}[4]{%
  \noindent
  \begin{minipage}[t]{0.485\linewidth}\raggedright\textbf{NU Grid:} #3\end{minipage}\hfill
  \begin{minipage}[t]{0.485\linewidth}\raggedright\textbf{U Grid:} #1\end{minipage}\par
  \vspace{0.15\baselineskip}
  \begin{minipage}[t]{0.485\linewidth}\raggedright\emph{Ops.:} #4\end{minipage}\hfill
  \begin{minipage}[t]{0.485\linewidth}\raggedright\emph{Ops.:} #2\end{minipage}%
}
\newcommand{\AlgoStepComm}[2]{%
  \item[\textbf{#1.}] \OneAlgoBox{\AlgoBoxTitle{#2}}%
}
\newcommand{\AlgoStepOne}[4]{%
  \item[\textbf{#1.}] \OneAlgoBox{%
    \AlgoBoxHead{#2}{#3}
    {\centering \emph{Ops.:} #4\par}%
  }%
}
\newcommand{\AlgoStepTwo}[7]{%
  \item[\textbf{#1.}] \OneAlgoBox{%
    \AlgoBoxHead{#2}{#3}
    \AlgoUniNonUniOps{#4}{#5}{#6}{#7}%
  }%
}
\vspace{0.25\baselineskip}

\AlgoSection{Initialization}
\begin{enumerate}[itemsep=0.05\baselineskip]
  \item[]
    \OneAlgoBox{%
      \AlgoBoxHead{Setup transforms:}{For each direction $\alpha\in\{x,y\}$}
      \AlgoUniNonUni
        {Compute/store analytical eigenvalues based on boundary conditions \cite{Schumann-and-Sweet-JCP-1988}; initialize batches of $\alpha$-direction FFTs.}
        {Symmetrize $\boldsymbol{T}_\alpha$ with the boundary conditions included; compute the eigendecomposition using \texttt{xSTEDC}/\texttt{xSYEVD}; construct $\boldsymbol{Q}_\alpha$ and $\boldsymbol{Q}_\alpha^{-1}$ using Eq.~\eqref{eqn:q-from-tildeq}; store $\boldsymbol{\Lambda}_\alpha$, $\boldsymbol{Q}_\alpha$, and $\boldsymbol{Q}_\alpha^{-1}$.}%
    }
\end{enumerate}

\vspace{0.25\baselineskip}
\AlgoSection{Solution}
\begin{enumerate}[itemsep=0.05\baselineskip]
  \AlgoStepTwo{1}{Forward $x$ transform:}{$n_y^x n_z^x$ lines of length $N_x$}{Batch of 1D $x$ FFT-based transforms.}{$n_y^x n_z^x\, O(N_x \log N_x)$}{GEMM with $\boldsymbol{Q}_x^{-1}$.}{$n_y^x n_z^x\, O(N_x^2)$}
  \AlgoStepComm{2}{Transpose $x\,\rightarrow y$\textnormal{: from ($N_x\times n_y^x\times n_z^x$) to ($n_x^y\times N_y\times n_z^y$) subdomains}}
  \AlgoStepTwo{3}{Forward $y$ transform:}{$n_x^y n_z^y$ lines of length $N_y$}{Batch of 1D $y$ FFT-based transforms.}{$n_x^y n_z^y\, O(N_y \log N_y)$}{GEMM with $\boldsymbol{Q}_y^{-1}$.}{$n_x^y n_z^y\, O(N_y^2)$}
  \AlgoStepComm{4}{Transpose $y\,\rightarrow z$\textnormal{: from ($n_x^y\times N_y\times n_z^y$) to ($n_x^z\times n_y^z\times N_z$) subdomains}}
  \AlgoStepOne{5}{Tridiagonal solve:}{$n_x^z n_y^z$ systems of length $N_z$ with TDMA}{$n_x^z n_y^z\, O(N_z)$}
  \AlgoStepComm{6}{Transpose $z\,\rightarrow y$\textnormal{: from ($n_x^z\times n_y^z\times N_z$) to ($n_x^y\times N_y\times n_z^y$) subdomains}}
  \AlgoStepTwo{7}{Inverse $y$ transform:}{$n_x^y n_z^y$ lines of length $N_y$}{Batch of 1D $y$ inverse FFT-based transforms.}{$n_x^y n_z^y\, O(N_y \log N_y)$}{GEMM with $\boldsymbol{Q}_y$.}{$n_x^y n_z^y\, O(N_y^2)$}
  \AlgoStepComm{8}{Transpose $y \rightarrow x$\textnormal{: from ($n_x^y\times N_y\times n_z^y$) to ($N_x\times n_y^x\times n_z^x$) subdomains}}
  \AlgoStepTwo{9}{Inverse $x$ transform:}{$n_y^x n_z^x$ lines of length $N_x$}{Batch of 1D $x$ inverse FFT-based transforms.}{$n_y^x n_z^x\, O(N_x \log N_x)$}{GEMM with $\boldsymbol{Q}_x$.}{$n_y^x n_z^x\, O(N_x^2)$}
\end{enumerate}

\end{mdframed}

\end{algorithm}

When GEMM-based transforms are used along both diagonalized directions, the arithmetic work per solve scales as $n_y^x n_z^x\,O(N_x^2)+n_x^y n_z^y\,O(N_y^2)+n_x^z n_y^z\,O(N_z)$ \cite{Farnell-JCP-1980,Vitoshkin-and-Gelfgat-CCP-2013,Krasnov-et-al-JCP-2023}. For a uniform direction $\alpha\in\{x,y\}$, an FFT-based transform reduces the per-line cost from $O(N_\alpha^2)$ to $O(N_\alpha\log N_\alpha)$. The linear $z$-direction TDMA contribution is unchanged.

Despite their higher asymptotic complexity, GEMM-based transforms have high arithmetic intensity and map efficiently to optimized BLAS kernels on modern GPUs. FFT-based transforms generally have lower arithmetic intensity and require more local data movement. Both variants follow the same sequence of global pencil transposes, so the more compute-intensive GEMM variants can better amortize this shared communication pattern at scale. The performance results below quantify this trade-off while accounting for the additional flexibility provided by non-uniform grids.

\FloatBarrier
\section{Results}
\label{sec:results}

\subsection{Verification and validation}

We first verify the direct elliptic solver. For every supported grid and combination of boundary conditions, the algebraic residual $\boldsymbol{f}-\boldsymbol{T}\boldsymbol{\phi}$ is reduced to round-off at every grid point. Applying the pressure-projection step to a white-noise velocity field likewise produces a discretely divergence-free field to machine precision (not shown). The flow cases below then validate the implicit-diffusion path and the complete Navier--Stokes method.

We validate the complete flow solver using three cases on stretched meshes: a three-dimensional lid-driven cavity, a two-dimensional lid-driven cavity in the Stokes limit, and a turbulent square duct. The two-dimensional Stokes case separately isolates the implicit Helmholtz treatment of diffusion. Grid points in each stretched direction are distributed using a conventional hyperbolic-tangent mapping with clustering at both ends \cite{Orlandi-2012}. For a direction of length $L$, the mapping from a uniform computational coordinate $\xi\in[0,1]$ to the physical coordinate $x\in[0,L]$ is
\begin{equation}
  x(\xi)=\frac{L}{2}\left(
    1+
    \frac{\tanh\left[\alpha\left(\xi-\tfrac{1}{2}\right)\right]}
         {\tanh\left(\tfrac{\alpha}{2}\right)}
  \right),
  \label{eq:tanh_stretching}
\end{equation}
where $\alpha\geq 0$ controls the clustering intensity near the boundaries\footnote{The limit $\alpha\to0$ gives a uniform grid. For a sufficiently large number of grid points, the maximum-to-minimum spacing ratio is well approximated by $\Delta_{\max}/\Delta_{\min}\approx\cosh^2(\alpha/2)$.}.

\begin{table}
  \centering
   \caption{Physical and computational parameters for the validation cases. $L_x,L_y,L_z$, $N_x,N_y,N_z$, and $\alpha_x,\alpha_y,\alpha_z$ denote the domain lengths, numbers of grid points, and stretching parameters, respectively (see Eq.~\eqref{eq:tanh_stretching}).}
   \label{tbl:validation-cases}
   \begin{tabular}{ l c c c c }
   \hline\hline
    Case                  & $L_x/h,L_y/h,L_z/h$ & $N_x,N_y,N_z$ & $\alpha_x,\alpha_y,\alpha_z$ & $\mathrm{Re} = U h/\nu$ \\
   \hline\hline
    3D lid-driven cavity  & $1,1,1$       & $100,100,100$ & $1,1,1$                      & $1000$ \\
   \hline
    2D Stokes cavity
                          & $1,1,\text{--}$ & $64,64,\text{--}$ & $1,1,\text{--}$          & $0.001$ \\
   \hline
    Turbulent square duct & $10,1,1$      & $512,100,100$ & $0,2,2$                      & $4410$ \\
   \hline
   \hline
  \end{tabular}
  \end{table}

\subsubsection*{Lid-driven cavity flows}

We validate the three-dimensional lid-driven cavity in a cubic domain of side $h$, with $(x,y,z)\in[-h/2,h/2]^3$, against the reference DNS data of Ku et al.~\cite{Ku-et-al-JCP-1987}. The physical and computational parameters are listed in Table~\ref{tbl:validation-cases}, and the Reynolds number is defined as $\mathrm{Re}=Uh/\nu$. No-slip boundary conditions are imposed at all walls. The lid at $y=h/2$ moves with velocity $\mathbf{u}(x,h/2,z)=(U,0,0)$, while the remaining walls are stationary. Accordingly, the pressure Poisson equation uses homogeneous Neumann boundary conditions at all walls.

The two-dimensional Stokes-limit cavity at $\mathrm{Re}=0.001$ validates the implicit-diffusion implementation. We compute its reference solution using a vorticity--streamfunction finite-difference formulation \cite{Hinch-2020}. The stretching parameters match the 3D case, while a $64\times64$ grid is sufficient to resolve the flow. This diffusion-dominated case provides a sensitive test of the unsplit Helmholtz solve in Eq.~\eqref{eqn:helmholtz_ns}, because the splitting terms introduced by the approximate-factorization alternatives discussed in Section~\ref{sec:Navier--Stokes-solver} can become appreciable.

Fig.~\ref{fig:validation}(\textit{a}) compares steady-state centerline velocity profiles with the DNS data of Ku et al.~\cite{Ku-et-al-JCP-1987} and the Stokes-limit reference solution. The results agree closely with their respective references.

\subsubsection*{Turbulent square duct flow}

We next validate the solver against the reference DNS of Gavrilakis \cite{Gavrilakis-JFM-1992} for turbulent flow through a square duct at friction Reynolds number $\mathrm{Re}_\tau\approx150$. The duct has side $h$ and streamwise length $L_x=10h$, with $(y,z)\in[-h/2,h/2]^2$. The flow is periodic along $x$ and satisfies no-slip boundary conditions at the four duct walls. Accordingly, the pressure Poisson equation uses periodic boundary conditions along $x$ and homogeneous Neumann conditions in the cross-stream directions. We drive the flow with a spatially uniform pressure-gradient forcing $G$, adjusted at every time step to maintain the bulk velocity $U$, giving the bulk Reynolds number $\mathrm{Re}=Uh/\nu=4410$. The grid has $\Delta x^+\approx5.79$, while the cross-stream spacing ranges from $\Delta y^+_{\min}=\Delta z^+_{\min}=1.66$ near the walls to $\Delta y^+_{\max}=\Delta z^+_{\max}=3.89$ at the duct center. Here, the superscript $+$ denotes viscous scaling by $\nu/u_\tau$.

\begin{figure}[H]
  \centering
  \includegraphics[height=0.49\textwidth]{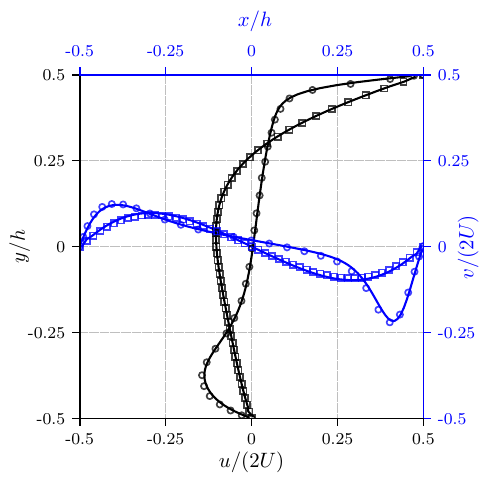}\hfill
  \includegraphics[height=0.44\textwidth]{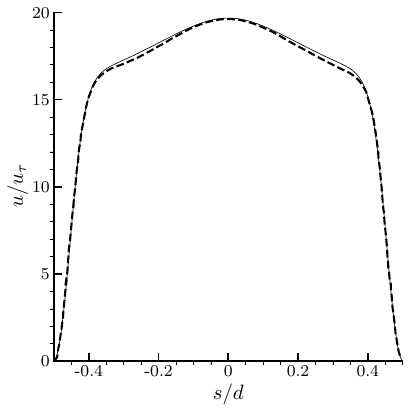}
  \put(-460pt,210pt){\small(\textit{a})}
  \put(-205pt,210pt){\small(\textit{b})}
  \caption{Validation test cases. Solid lines denote the present results. (\textit{a}) Lid-driven cavities: velocity profiles along the centerlines of the 3D cavity at $\mathrm{Re}=1000$ (center plane $z=0$) and the 2D cavity in the Stokes limit ($\mathrm{Re}\approx0$). Circles denote the reference DNS data of Ku et al.~\cite{Ku-et-al-JCP-1987}, and squares denote the reference vorticity--streamfunction solution. The blue profiles represent the wall-normal velocity component $v$ along $x$, and the black profiles represent the streamwise velocity component $u$ along $y$. (\textit{b}) Turbulent square duct at $\mathrm{Re}=4410$: mean streamwise velocity along the cross-section diagonal, where $s=\sqrt{2}\,y=\sqrt{2}\,z$ and $d=\sqrt{2}\,h$ is the diagonal length. The dashed line denotes the reference DNS data of Gavrilakis \cite{Gavrilakis-JFM-1992}.}
  \label{fig:validation}
\end{figure}

The time-averaged forcing gives $u_\tau=\sqrt{\langle G\rangle h/4}$ and $\mathrm{Re}_\tau=u_\tau h/(2\nu)\approx148.3$, consistent with the reference value. Statistics are averaged over $800$ snapshots spanning $100$ flow-through times $L_x/U$. Fig.~\ref{fig:validation}(\textit{b}) compares the mean streamwise velocity along the duct diagonal with the reference DNS; the computed profile agrees closely with the reference data.

\subsection{Computational performance}

We now assess the cost and scalability of the present method on CPUs and GPUs. The performance tests use fully explicit momentum diffusion; assessment of the selectable implicit-diffusion options is left for future work. Detailed timings focus on the pressure Poisson solve because it contains the global transposes and directional transforms that distinguish the present approach. The remaining flow-solver workload consists mainly of local finite-difference operations and nearest-neighbor halo exchanges, whose behavior on large GPU systems has been examined by Diez et al.~\cite{Diez-et-al-CPC-2025}. To place the Poisson results in their application context, we also report timings for the complete Navier--Stokes solver.

All performance tests use the pressure boundary conditions of the turbulent square-duct case: periodic along $x$ and homogeneous Neumann along $y$ and $z$. On a uniform grid, these conditions lead to a Fourier transform along $x$ and a fast cosine transform (FCT) along $y$\footnote{The FCT is evaluated using FFTs following Makhoul \cite{Makhoul-IEEE-1980}.}. We compare these FFT-based transforms with their GEMM-based counterparts on a uniform mesh of the same size. These comparisons quantify the additional transform cost and allow us to estimate the reduction in grid-point count required for mesh stretching to offset it.

We first examine CPU performance against reference solvers and then assess single- and multi-GPU performance. All computations use double precision. The CPU tests were compiled with the GNU toolchain (\texttt{gfortran}~\texttt{13.3.0}) and linked against \texttt{OpenMPI}~\texttt{5.0.3}, \texttt{FFTW}~\texttt{3.3.10}, \texttt{OpenBLAS}~\texttt{0.3.27}, and \texttt{hypre}~\texttt{3.0.0}. The GPU tests used the \texttt{NVIDIA HPC SDK}~\texttt{25.7}. Unless stated otherwise, reported timings are arithmetic means of the $20$ fastest post-warm-up samples.

\subsubsection{CPU performance against reference solvers}

We compare five strategies for solving the Poisson equation:
\begin{enumerate}[itemsep=0em,topsep=0.5em]
  \item Present method: the four FFT/GEMM transform combinations defined below;
  \item \underline{3D PFMG}: stand-alone parallel semicoarsening multigrid (PFMG) applied to the full three-dimensional problem using the structured-grid solvers from \texttt{hypre} \cite{Falgout-et-al-2002};
  \item \underline{3D Krylov--MG}: \texttt{BiCGSTAB} applied to the full three-dimensional problem, with one V-cycle of PFMG per preconditioner application, denoted \texttt{BiCGSTAB+PFMG(1)} \cite{Falgout-et-al-2002};
  \item \underline{2D Krylov--MG + FFT}: FFT along $x$, followed by \texttt{BiCGSTAB+PFMG(1)} on the reduced 2D systems;
  \item \underline{FFT+\texttt{BLKTRI}}: FFT-based diagonalization along $x$, followed by direct two-dimensional solves using the \texttt{FISHPACK} block-cyclic-reduction routine \texttt{BLKTRI} \cite{FISHPACK,Swarztrauber-and-Sweet-JCAM-1989,Orlandi-and-Pirozzoli-JoT-2020}.
\end{enumerate}
The last two strategies require uniform spacing along $x$. For every iterative test, the relative residual tolerance is $10^{-4}$, and the maximum iteration count is set high enough to reach this tolerance. The multigrid configurations use the implementation in the DNS solver \texttt{SNaC} \cite{Costa-CPC-2022}\footnote{For 3D~PFMG and 3D~Krylov--MG, the Poisson equation is multiplied by the local cell volume $\Delta x_i\Delta y_j\Delta z_k$; for 2D Krylov--MG + FFT, the reduced equation is multiplied by the local cell area $\Delta y_j\Delta z_k$. These scalings give the symmetric second-order finite-volume form of the Laplace operator and improve convergence on stretched meshes.}.

For the present method, the first and second letters identify the transforms applied along $x$ and $y$, respectively. Here, \texttt{F} denotes a Fourier-based transform and \texttt{G} a numerical eigenbasis transform evaluated with GEMM:
\begin{itemize}[itemsep=0pt,topsep=0.5em]
  \item[-] \texttt{FF}: FFT along $x$ and FCT along $y$;
  \item[-] \texttt{GF}: GEMM along $x$ and FCT along $y$;
  \item[-] \texttt{FG}: FFT along $x$ and GEMM along $y$;
  \item[-] \texttt{GG}: GEMM along both $x$ and $y$.
\end{itemize}

For the multigrid comparisons, we use three meshes generated with $\alpha\in\{0,2,4\}$ in Eq.~\eqref{eq:tanh_stretching}, denoted \texttt{GR0}, \texttt{GR2}, and \texttt{GR4}. The stand-alone PFMG and 3D Krylov--MG cases are stretched along all three directions, whereas the 2D Krylov--MG + FFT cases retain uniform spacing along $x$ and are stretched only along $y$ and $z$. At fixed grid dimensions, the cost of the direct methods is insensitive to mesh grading; we therefore report one timing for each direct-solver configuration.

\subsubsection*{Single-core performance}

We first consider a single-core CPU benchmark on a $128^3$ cubic grid. Relative to \texttt{FF}, the mixed cases \texttt{FG} and \texttt{GF} are about $1.3\times$ slower, while the fully GEMM-based case \texttt{GG} is $1.7\times$ slower. Despite this additional transform cost, all GEMM-based variants remain $3.9$--$5.2\times$ faster than FFT+\texttt{BLKTRI}.

\begin{figure}
    \centering
    \includegraphics[width=0.99\linewidth]{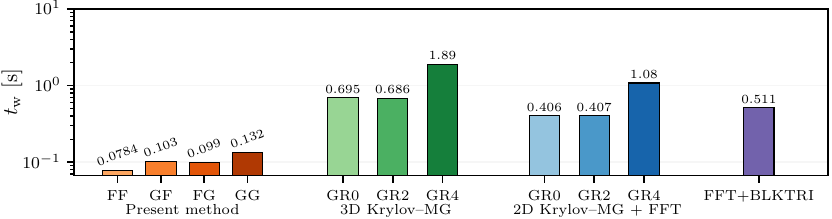}
    \caption{Single-core CPU performance for the Poisson solve on a $128^3$ cubic grid with the same boundary conditions as the square-duct case. The vertical axis reports the mean wall-clock time of the $20$ fastest solves per case. The 3D Krylov--MG and 2D Krylov--MG + FFT bars use \texttt{BiCGSTAB+PFMG(1)}; see the main text for the remaining case definitions. In the corresponding complete flow-solver measurements, the pressure solution accounts for about $50$--$63\%$ of the overall runtime for the present method, $87\%$ for FFT+\texttt{BLKTRI}, and $76$--$94\%$ for the Krylov--MG cases shown here.}
    \label{fig:single-core-performance}
\end{figure}

Fig.~\ref{fig:single-core-performance} summarizes the single-core comparison, showing the present FFT/GEMM variants and FFT+\texttt{BLKTRI} alongside the 3D Krylov--MG and 2D Krylov--MG + FFT configurations. Stand-alone 3D PFMG is omitted because its cost increases much more rapidly with mesh grading: its Poisson-solve time is within $10\%$ of 3D Krylov--MG on \texttt{GR0}, but reaches $6.5\times$ that of 3D Krylov--MG on \texttt{GR4}. Accordingly, 3D Krylov--MG is used in the figure, rather than stand-alone 3D PFMG, as the iterative reference for the full three-dimensional problem.

Embedding one PFMG V-cycle within \texttt{BiCGSTAB} allows the outer Krylov iteration to correct residual components that the multigrid hierarchy reduces inefficiently, thereby moderating the sensitivity to stretching-induced anisotropy.

For 3D Krylov--MG (stretched in all directions\footnote{For the present grid sizes, the grading ratios in each stretched direction are $\Delta_{\max}/\Delta_{\min}\approx 2.35$ and $13.7$ for $\alpha=2$ and $4$.}), the Poisson solve takes $t_\mathrm{w}=0.695\,\mathrm{s}$ on \texttt{GR0} and remains essentially unchanged at $0.686\,\mathrm{s}$ on \texttt{GR2}, before rising to $1.895\,\mathrm{s}$ on \texttt{GR4} ($2.7\times$ the \texttt{GR0} value). Applying FFT-based diagonalization along $x$ reduces these times to $0.406$, $0.407$, and $1.081\,\mathrm{s}$, respectively, a speedup of $1.68$--$1.75\times$ over the full 3D configuration. This deterioration on strongly graded meshes is consistent with the induced anisotropy weakening the multigrid ``frequency separation'' assumption, namely the smoothing of oscillatory error followed by coarse-grid correction of smooth components \cite{Trottenberg-et-al-2001}. On \texttt{GR4}, the 3D Krylov--MG and 2D Krylov--MG + FFT solves are about $14.3\times$ and $10.9\times$ slower than their compatible direct counterparts, \texttt{GG} and \texttt{FG}.

These are not equal-accuracy comparisons: at the prescribed $10^{-4}$ iterative tolerance, the maximum recorded post-projection divergences for the \texttt{GR4} full 3D and FFT-reduced tests are $1.52\times10^{-5}$ and $5.87\times10^{-6}$, compared with $O(10^{-9})$ for the direct method on the same grid.

\begin{figure}
  \centering
  \includegraphics[width=0.99\linewidth]{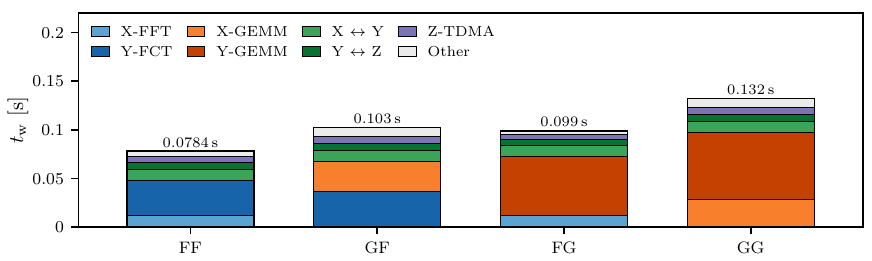}
  \caption{Operation-level breakdown of the single-core CPU time per Poisson solve on a $128^3$ cubic grid for the $x$/$y$ transform combinations \texttt{FF}/\texttt{GF}/\texttt{FG}/\texttt{GG}: FFTs/GEMMs along $x$ and $y$ (blues/reds; forward and inverse combined), collective pencil transposes (greens; forward and inverse combined), and the TDMA step along $z$ (purples). ``Other'' denotes minor contributions from local array copies (light gray). Across these four cases, the pressure solution represents about $50$--$63\%$ of the overall Navier--Stokes runtime.}
  \label{fig:single-core-performance-breakdown}
\end{figure}

Fig.~\ref{fig:single-core-performance-breakdown} decomposes the timings for the four FFT/GEMM transform combinations. Differences in overall wall-clock time are primarily driven by the $x$- and $y$-direction transform costs. In all cases, the dominant contribution comes from the $y$ direction. Local transposes are required to pack and unpack the data, while the FCT imposed by the boundary conditions accounts for roughly $35$--$46\%$ of the solve time in \texttt{FF} and \texttt{GF}. The $x$-direction FFT contributes about $12$--$15\%$ in \texttt{FF} and \texttt{FG}. Replacing the $x$ transform with GEMM increases its share to approximately $22$--$30\%$ in \texttt{GF} and \texttt{GG}. When GEMM is used along $y$, the transform and its associated local packing and unpacking account for approximately $52$--$62\%$ of the solve time in \texttt{FG} and \texttt{GG}. The $z$-direction TDMA step remains a secondary contribution, consistent with its linear complexity. The global-transpose code path also incurs a small overhead on one core because it is not specialized for the single-task case.

\subsubsection*{Many-core performance}

Strong- and weak-scaling tests were performed on the AMD Rome thin partition of the Snellius supercomputer (SURF, Amsterdam), where each node comprises $128$ CPU cores. The scaling tests compare the present method with FFT+\texttt{BLKTRI} and stand-alone PFMG. FFT+\texttt{BLKTRI} is the fastest alternative in the single-core comparison, while PFMG provides an iterative reference that supports non-uniform grids in all directions. For strong scaling, we use stand-alone PFMG on the uniform \texttt{GR0} mesh rather than the \texttt{BiCGSTAB+PFMG(1)} configuration, since their \texttt{GR0} performance differs little. Because mesh stretching penalizes this PFMG configuration, the uniform-grid series provides an optimistic reference for its time-to-solution.

For the strong scaling tests, we consider a cubic domain of size $N_x\times N_y\times N_z = 1024^3$ and vary the total number of CPU cores from $N_\mathrm{CPU}= 128$ to $8192$. The domain is partitioned among tasks in $x$-aligned pencils as described in Section~\ref{sec:implementation}, with a two-dimensional processor grid of size $P_1\times P_2 = (N_\mathrm{CPU}/64)\times 64$. The exception is the FFT+\texttt{BLKTRI} solver, which requires a one-dimensional slab decomposition ($N_\mathrm{CPU}\times1$), and, therefore, is limited\footnote{Note that the scale of these tests could be extended by adding shared-memory parallelism (using, e.g., \texttt{OpenMP}). We deem exploring hybrid (e.g., MPI+OpenMP) implementations beyond the scope of the present study.} to $N_\mathrm{CPU}\le N_y=1024$.

\begin{figure}
  \centering
  \begin{minipage}{0.49\linewidth}
      \centering
      \includegraphics[width=\linewidth]{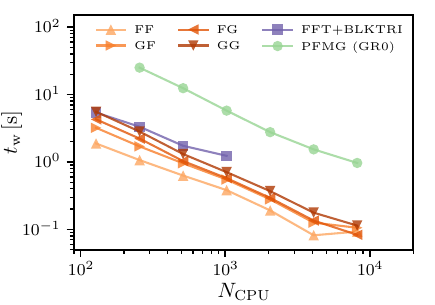}
  \end{minipage}\hfill
  \begin{minipage}{0.49\linewidth}
      \centering
      \includegraphics[width=\linewidth]{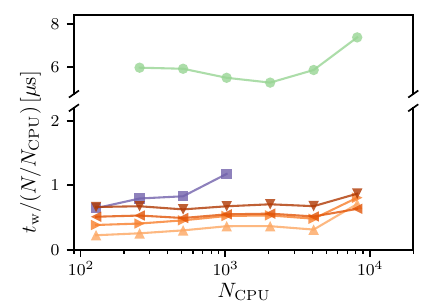}
  \end{minipage}
  \put(-470pt,80pt){\small(\textit{a})}
  \put(-225pt,80pt){\small(\textit{b})}
  \caption{Strong scaling of the Poisson solve on a $1024^3$ cubic grid. (\textit{a}) Wall-clock time per solve $t_\mathrm{w}$ as a function of the number of CPU cores $N_\mathrm{CPU}$ for the present method (orange curves) with different $x$/$y$ transform combinations (\texttt{FF}/\texttt{GF}/\texttt{FG}/\texttt{GG}), compared against FFT+\texttt{BLKTRI} (purple) and stand-alone PFMG on \texttt{GR0} (green). (\textit{b}) Task-local wall-clock time per grid cell $t_\mathrm{w}/(N/N_\mathrm{CPU})$ in microseconds, where $N=N_xN_yN_z$, as a function of $N_\mathrm{CPU}$; the $y$-axis is broken to accommodate the significantly larger PFMG values. Over the measured core counts, the pressure solution accounts for approximately $51$--$76\%$ of the overall Navier--Stokes runtime for the present variants, $55$--$63\%$ for FFT+\texttt{BLKTRI}, and $93\%$ for PFMG.}
  \label{fig:strong-scaling-all-panels}
\end{figure}

Fig.~\ref{fig:strong-scaling-all-panels} shows that all methods initially benefit from strong scaling over the considered range. Panel (\textit{a}) reports the nominal wall-clock time in seconds, and panel (\textit{b}) reports the wall-clock time per local grid cell, for which ideal scaling is horizontal. For the present solver, strong scaling generally improves as Fourier-based transforms are replaced by GEMM-based transforms, and deterioration remains small up to $N_\mathrm{CPU}=4096$, with \texttt{FF} showing the most variation and \texttt{GG} the least. The more compute-intensive GEMM transforms better amortize communication and transpose overheads at high core counts. Between $4096$ and $8192$ cores, the corresponding two-point parallel efficiencies are approximately $44\%$ for \texttt{FF} and $78\%$ for \texttt{GG}. FFT+\texttt{BLKTRI} scales well up to its slab-decomposition limit ($N_\mathrm{CPU}=1024$), where its performance deteriorates substantially. This case is $1.3$--$2.1\times$ slower than the mixed \texttt{FG} variant, which likewise supports uniform spacing along $x$ and non-uniform spacing along $y$ and $z$. Finally, stand-alone PFMG on a uniform mesh remains about one order of magnitude more expensive. This gap would widen on stretched grids or under stricter iteration tolerances chosen to match the accuracy of the direct method. The same trends are visible in the complete Navier--Stokes solver, since pressure accounts for $51$--$76\%$ of the present-method runtime and rises to $69$--$76\%$ at $8192$ cores; with PFMG, its contribution remains about $93\%$.

\begin{figure}
  \centering
  \includegraphics[width=0.99\linewidth]{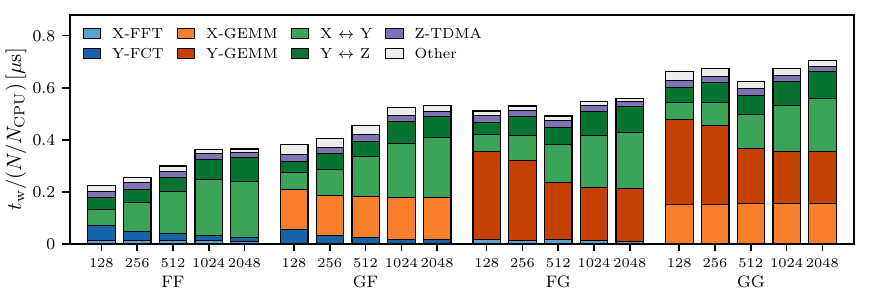}
  \caption{Operation-level breakdown of strong-scaling performance for the present solver on a $1024^3$ cubic grid, shown as normalized wall-clock time per grid cell $t_\mathrm{w}/(N/N_\mathrm{CPU})$ in microseconds ($N=N_xN_yN_z$) for the $x$/$y$ transform combinations \texttt{FF}/\texttt{GF}/\texttt{FG}/\texttt{GG}. The contributions are FFT/GEMM transforms along $x$ and $y$ (blues/reds; forward and inverse combined), collective pencil transposes (greens; forward and inverse combined), the TDMA step along $z$ (purples), and local array copies (``Other,'' light gray). In these runs, the pressure solution represents approximately $51$--$76\%$ of the overall Navier--Stokes runtime.}
  \label{fig:strong-scaling-eigen-breakdown-multicore-normalized}
\end{figure}

Fig.~\ref{fig:strong-scaling-eigen-breakdown-multicore-normalized} shows how computation and global transposes contribute to the normalized Poisson-solve time up to $N_\mathrm{CPU}=2048$. For the purely FFT-based case \texttt{FF}, global transposes already account for almost half of the solve at $N_\mathrm{CPU}=128$ and increase to about $84\%$ at $N_\mathrm{CPU}=2048$. In contrast, \texttt{GG} remains more compute-rich: at $N_\mathrm{CPU}=128$, about $81\%$ of the solve is spent in transforms, TDMA, and other local work, with only $19\%$ in global transposes; at $N_\mathrm{CPU}=2048$, the transpose fraction increases to about $44\%$. Consequently, the performance gap between \texttt{FF} and \texttt{GG} narrows from about $3\times$ at $128$ cores to $1.9\times$ at $2048$.

From a practical standpoint, the increased cost of \texttt{GG} must be offset by the reduction in total grid size enabled by stretched meshes. Based on the measured CPU timings, if an FFT-based reference run requires uniform spacing comparable to the smallest grid spacing, reducing the total number of cells by roughly a factor of $2$--$3$ would make \texttt{GG} competitive in time-to-solution. Grid design can also reduce the number of time steps. For a representative pipe-flow calculation, Pirozzoli and Orlandi \cite{Pirozzoli-and-Orlandi-JCP-2021} reduced the radial grid from $512$ points with conventional hyperbolic-tangent stretching to $270$ points with their natural mapping, while increasing the maximum inner-scaled time step from $\Delta t^+=0.12$ to $0.22$ without materially changing the primary statistics. Thus, a well-designed non-uniform mesh can save both cells and time steps.

Finally, we assess weak scaling by increasing the core count from $N_\mathrm{CPU}=128$ to $2048$ while keeping the number of grid cells per task fixed. The baseline case has a $768^3$ grid and processor grid\footnote{Because FFT+\texttt{BLKTRI} requires a slab decomposition, its one-dimensional decomposition grows linearly with $N_\mathrm{CPU}$ while $N_x$ increases proportionally.} $P_1^0\times P_2=2\times64$. We keep $P_2=64$ fixed and set $P_1=N_\mathrm{CPU}/P_2$. The global resolution increases only along $x$, according to $N_x=N_x^0P_1/P_1^0$, with $N_x^0=N_y^0=N_z^0=768$, while $N_y=N_y^0$ and $N_z=N_z^0$. The resulting $x$-pencil dimensions are $(N_x,N_y^0/P_1,N_z^0/P_2)$, whose product is constant. The largest case is therefore $12288\times768\times768$, with a $16{:}1$ grid-size aspect ratio. This setup isolates the different growth rates of FFT- and GEMM-based transforms along the expanding direction.

\begin{figure}
  \centering
  \includegraphics[width=0.49\linewidth]{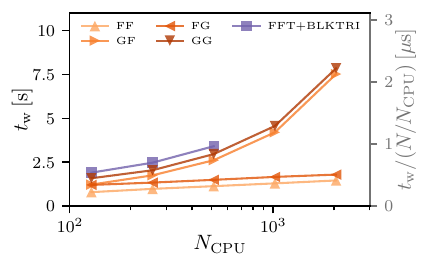}\hfill
  \includegraphics[width=0.49\linewidth]{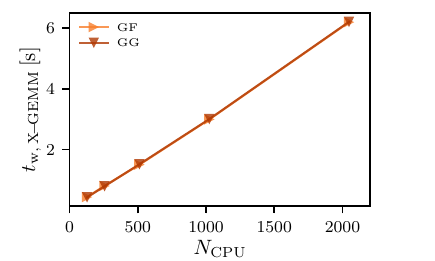}
  \put(-470pt,130pt){\small(\textit{a})}
  \put(-225pt,130pt){\small(\textit{b})}
  \caption{(\textit{a}) Wall-clock time per Poisson solve versus the core count $N_\mathrm{CPU}$, in weak scaling tests with increasing $N_x$ while keeping the local problem size fixed, for the present method (orange curves) with different $x$/$y$ transform combinations, compared against FFT+\texttt{BLKTRI} (purple). The left axis shows the nominal Poisson solver wall-clock time $t_\mathrm{w}$ in seconds, while the right one shows the normalized wall-clock time per local grid cell $t_\mathrm{w}/(N/N_\mathrm{CPU})$ in microseconds. (\textit{b}) Wall-clock time of the {X-GEMM} operation for the cases using this operation along $x$: \texttt{GF} and \texttt{GG}, showing a linear trend. Across the weak-scaling sequence, the pressure solution accounts for approximately $50$--$88\%$ of the overall Navier--Stokes runtime for the present variants and $55$--$69\%$ for FFT+\texttt{BLKTRI}.}
  \label{fig:weak-scaling-inc-x-dual-yaxis}
\end{figure}

Fig.~\ref{fig:weak-scaling-inc-x-dual-yaxis}(\textit{a}) presents the wall-clock time per Poisson solve, with stand-alone PFMG omitted because of its much higher cost. FFT+\texttt{BLKTRI} is relatively competitive at low core counts, but its solve time grows noticeably under weak scaling as the communication footprint of the slab-based formulation increases.
For the present approach, the main distinction is between cases that use a Fourier-based transform along the growing direction $x$ (\texttt{FF} and \texttt{FG}) and those that use a GEMM-based transform (\texttt{GF} and \texttt{GG}). The per-line costs are $O(N_x\log N_x)$ for FFTs and $O(N_x^2)$ for GEMM. Multiplying by the number of lines per task, $n_y^x n_z^x$, gives $t_\mathrm{w}\sim\log(N_\mathrm{CPU})$ for FFT-based transforms and $t_\mathrm{w}\sim N_\mathrm{CPU}$ for GEMM-based transforms once the $x$ operation dominates. The linear GEMM growth is demonstrated in panel~(\textit{b}) of Fig.~\ref{fig:weak-scaling-inc-x-dual-yaxis}.
\begin{figure}
  \centering
  \includegraphics[width=0.99\linewidth]{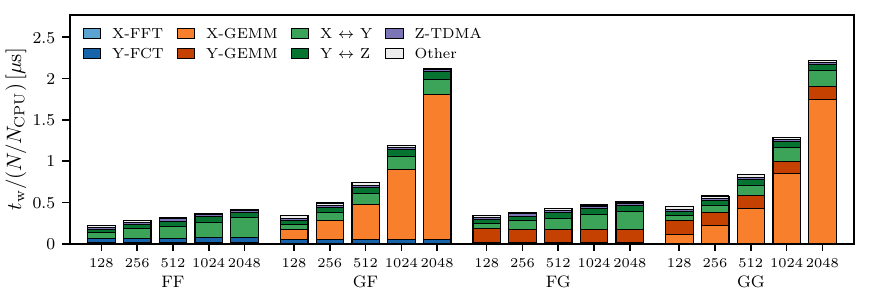}\hfill
  \caption{Same as Fig.~\ref{fig:strong-scaling-eigen-breakdown-multicore-normalized}, but for the present weak scaling tests with increasing $N_x$ and $N_\mathrm{CPU}$ while keeping the local problem size fixed. In these runs, the pressure solution represents approximately $50$--$88\%$ of the overall Navier--Stokes runtime.}
  \label{fig:weak-scaling-eigen-breakdown-multicore-normalized}
\end{figure}

The Poisson timings in Fig.~\ref{fig:weak-scaling-inc-x-dual-yaxis}(\textit{a}) and their operation-level breakdown in Fig.~\ref{fig:weak-scaling-eigen-breakdown-multicore-normalized} reflect these trends. For the purely FFT-based case \texttt{FF}, the local Poisson time increases by $1.8\times$ from $128$ to $2048$ cores. The $x$-FFT contribution remains marginal at $5.8$--$7.3\%$, while the transpose contribution increases from $48.1\%$ to $71.5\%$. When GEMM is used along $x$, that transform rapidly becomes dominant: the Poisson time increases by $6.1\times$ for \texttt{GF} and $4.9\times$ for \texttt{GG}, with the $x$ transform accounting for $79$--$82\%$ of the solve at the largest core count. Even so, both cases remain faster than FFT+\texttt{BLKTRI}. This growing-direction cost also shapes the application performance: pressure accounts for as much as $88\%$ of the complete Navier--Stokes runtime across the sequence and increasingly governs it as the $x$-GEMM becomes dominant.

Asymptotically, the quadratic cost of the dense eigenbasis transforms, and eventually even the logarithmic FFT cost, could be overtaken by iterative methods with linear complexity in the number of unknowns. No crossover is observed over the tested range, and the measured performance gap remains substantial. Moreover, the iterative benchmarks use a relatively loose residual tolerance of $10^{-4}$; matching the accuracy of the direct solve would require a stricter tolerance and further increase their cost, particularly on stretched meshes. The accelerator considerations discussed in Section~\ref{sec:introduction} are expected to push the practical crossover farther out on GPU systems, where the present formulation retains optimized FFT and GEMM kernels.

\subsubsection{Performance on GPUs}
\smallskip

\subsubsection*{Single-GPU performance}
\begin{figure}[H]
  \centering
  \includegraphics[width=\linewidth]{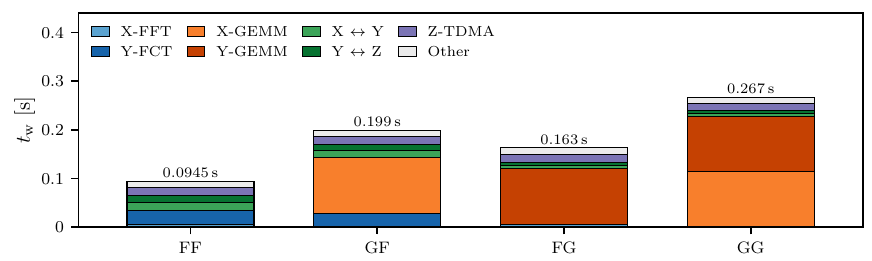}
  \caption{Same as Fig.~\ref{fig:single-core-performance-breakdown}, but for a $1024\times1024\times1024$ grid on a single GB200 superchip. Across the four variants, the pressure solution accounts for about $44\%$--$69\%$ of the overall Navier--Stokes runtime.}
  \label{fig:single-gpu-performance-breakdown}
\end{figure}

To assess single-GPU performance, we benchmark the four FFT/GEMM transform combinations using the same boundary-condition setup as above. The grid is increased to $1024^3$ to make effective use of an NVIDIA~GB200; the complete Navier--Stokes solver occupies about $60$--$70\%$ of its $192\,\mathrm{GB}$ memory. Fig.~\ref{fig:single-gpu-performance-breakdown} breaks down the wall-clock time per Poisson solve. The total ranges from $0.094\,\mathrm{s}$ for \texttt{FF} to $0.267\,\mathrm{s}$ for \texttt{GG} ($2.8\times$ higher), with \texttt{FG} and \texttt{GF} giving $0.163$ and $0.199\,\mathrm{s}$, respectively. These values correspond to $0.088$--$0.248\,\mathrm{ns}$ per grid cell. The $y$-direction operation and its local packing transposes already account for $61\%$ of the \texttt{FF} solve. Replacing the $y$ FCT with GEMM increases the arithmetic work in \texttt{FG} and \texttt{GG}, but reduces the transpose cost\footnote{FFT-based real-to-complex transforms use padded layouts with a leading dimension $N_x^\mathrm{FFT}=2(\lfloor N_x/2\rfloor+1)$ for the transpose input arrays, whereas the GEMM-based variants transpose real arrays of size $N_x$. The misalignment of memory in the padded layout can sometimes lead the underlying libraries to select less efficient routines (e.g., non-vectorized variants).}. For the complete Navier--Stokes solver, the corresponding increase is only about $1.8\times$ for the most expensive case, \texttt{GG}.

\subsubsection*{Multi-GPU performance}

We next assess the scalability of the Poisson solver on an NVIDIA GB200 NVL72 cluster through strong and weak scaling tests. In this system, each node comprises four GB200 GPUs connected with high bandwidth through NVLink/NVSwitch; across nodes, the system can either be configured as a single 72-GPU Multinode NVLink (MNNVL) domain, or as a conventional multi-node setup with lower bandwidth, relying on InfiniBand (here denoted no-MNNVL). The best-performing processor grid $P_1\times P_2$ and communication backend are selected using \texttt{cuDecomp}'s runtime autotuning capabilities.

\begin{figure}
  \centering
  \includegraphics[width=\linewidth]{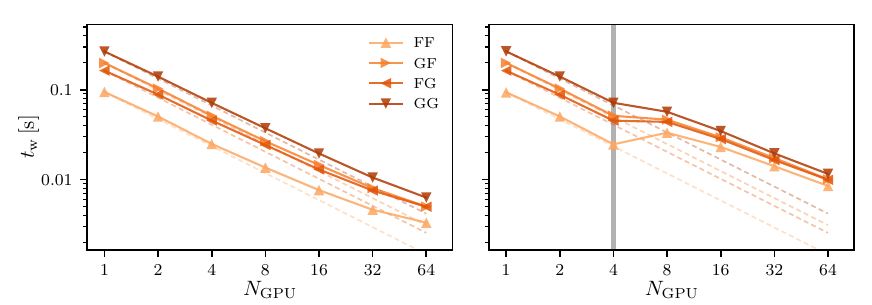}
  \put(-440pt,160pt){\small(\textit{a})}
  \put(-220pt,160pt){\small(\textit{b})}
  \caption{Strong scaling on GPUs for a $1024\times1024\times1024$ grid, comparing Multinode NVLink (\textit{a}) and a conventional InfiniBand interconnect (no-MNNVL; \textit{b}). The plots show the wall-clock time per Poisson solve in seconds versus the number of GPUs. Color-matched dashed lines denote ideal scaling, anchored at the first point. The gray vertical line at $N_\mathrm{GPU}=4$ for no-MNNVL marks the transition from intra-node to inter-node execution. Over the measured GPU counts, the pressure solution accounts for approximately $44$--$70\%$ of the overall Navier--Stokes runtime with MNNVL and $44$--$78\%$ without MNNVL.}
  \label{fig:gpu-strong-scaling-mnnvl-vs-nomnnvl}
\end{figure}

In the strong scaling tests, the global problem size is fixed to the previous single-GPU problem, a $1024^3$ grid, while the GPU count $N_\mathrm{GPU}$ is increased. Fig.~\ref{fig:gpu-strong-scaling-mnnvl-vs-nomnnvl} reports the wall-clock time per Poisson solve for the four transform variants \texttt{FF}/\texttt{GF}/\texttt{FG}/\texttt{GG}, together with color-matched ideal-scaling references anchored at the single-GPU case. In the no-MNNVL configuration, the vertical line at $N_\mathrm{GPU}=4$ marks the transition from intra-node scaling within a four-GPU NVLink/NVSwitch node to inter-node scaling over InfiniBand.

\FloatBarrier
On MNNVL, all four variants scale well to $64$ GPUs, with speedups of about $29$--$42\times$ and corresponding parallel efficiencies of about $45\%$--$66\%$. The more compute-rich GEMM-based variants (\texttt{GF} and \texttt{GG}) retain higher parallel efficiency. The MNNVL and no-MNNVL configurations perform similarly up to four GPUs within one node, but the no-MNNVL scaling efficiency decreases once execution spans multiple nodes. The drop at the intra-/inter-node boundary is most pronounced for \texttt{FF}, which slows from $N_\mathrm{GPU}=4$ to $8$. At larger scales, differences among the transform combinations decrease as communication becomes dominant. At $N_\mathrm{GPU}=64$, MNNVL is $1.8$--$2.6\times$ faster than no-MNNVL, depending on the $x$--$y$ transform pair. Pressure accounts for $51$--$67\%$ of the Navier--Stokes runtime with MNNVL at this scale and $69$--$72\%$ without it, so the inter-node communication penalty is amplified in the complete application.

In the weak scaling tests, we start from the same baseline case used in the strong scaling analysis and increase the global resolution by factors of $1$, $2$, $3$, and $4$ in each direction while keeping the local problem size constant. The resulting global grids have resolutions $1024^3$, $2048^3$, $3072^3$, and $4096^3$, with the GPU count increased proportionally to the problem size.

\begin{figure}
  \centering
  \includegraphics[width=\linewidth]{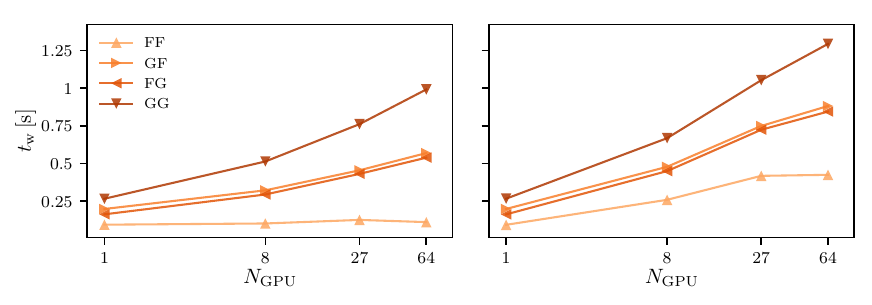}
  \caption{Weak scaling on GPUs, comparing Multinode NVLink (left) with a conventional InfiniBand interconnect (right). The panels report the same metrics as Fig.~\ref{fig:gpu-strong-scaling-mnnvl-vs-nomnnvl}. The global resolution is increased by a factor $\{1,2,3,4\}$ in each direction while increasing the GPU count by the corresponding factor $\{1,8,27,64\}$, so that the number of grid points per GPU is approximately constant. Across these runs, the pressure solution accounts for approximately $44$--$89\%$ of the overall Navier--Stokes runtime with MNNVL and $44$--$90\%$ without MNNVL.}\label{fig:gpu-weak-scaling-mnnvl-vs-nomnnvl}
\end{figure}
\FloatBarrier
Fig.~\ref{fig:gpu-weak-scaling-mnnvl-vs-nomnnvl} shows the Poisson solve time for this weak scaling test. The results are also strongly influenced by the communication substrate. The GEMM-heavy variants exhibit a stronger growth with scale (about $2.9$--$3.7\times$ on MNNVL and about $4.4$--$5.2\times$ on no-MNNVL). Hence, the MNNVL results more clearly expose the algorithmic weak-scaling trends, as the interconnect is fast enough to keep transpose overheads mild. The fully FFT-based \texttt{FF} case increases by only $1.2\times$ from $1$ to $64$ GPUs with MNNVL, compared with $4.5\times$ without MNNVL. Cases that use GEMM-based transforms grow substantially faster as the dense transforms become larger and increasingly dominate the solve. At $N_\mathrm{GPU}=27$ (global size $3072^3$), the \texttt{FF} case shows a mild local performance penalty because this size triggers a less bandwidth-efficient cuFFT execution path with an extra pass over the data. Since the FFT stages are memory-bandwidth-limited in our setup, this additional data movement increases their cost. At the largest weak-scaling point, pressure reaches $89\%$ of the Navier--Stokes runtime with MNNVL and $90\%$ without it. The transform-growth and interconnect effects seen in the Poisson solve can therefore dominate the application as well.

\FloatBarrier
\section{Conclusions and outlook}
\label{sec:conclusions-and-outlook}

We have presented a scalable finite-difference solver for the incompressible Navier--Stokes equations on three-dimensional Cartesian grids with non-uniform spacing. At its core is a direct tensor-product framework that treats both the pressure Poisson equation and the Helmholtz equations arising from implicit diffusion, without approximate factorization. For the full three-dimensional Poisson/Helmholtz problem, the operators in two directions are diagonalized, reducing the problem to independent tridiagonal systems in the third direction. A diagonal similarity transformation symmetrizes the non-uniform one-dimensional Laplace operators, allowing their eigenvectors to be computed efficiently. The resulting numerical eigenbasis transforms are evaluated with GEMMs, while uniform directions may instead use classical FFT-based eigenfunction expansions. This common formulation retains the pencil decomposition, collective transposes, and tridiagonal solution machinery of established FFT-based solvers while allowing FFTs and GEMMs to be selected independently in each diagonalized direction. The resulting solver, \texttt{CaNS-EIGEN}, is freely available and open-source.

On CPUs, the present method achieved the lowest Poisson-solve time among the tested alternatives. The compute-rich GEMM variants also showed systematically higher parallel efficiency in strong scaling because they amortized communication more effectively. Their greater arithmetic cost becomes apparent in weak scaling, however: an FFT applied along the growing direction leads to only mild runtime growth, whereas a GEMM eventually reflects the quadratic per-line cost of the dense transform. Since the pressure solution accounts for $51$--$76\%$ of the Navier--Stokes runtime in CPU strong scaling and up to $88\%$ in weak scaling, these pressure-solver trends strongly influence the performance of the complete application.

The GPU results show the same balance between computation and communication. On a single GPU superchip, replacing both FFT-based transforms with GEMM-based transforms made the Poisson solve $2.8\times$ more expensive, but increased the cost of a complete Navier--Stokes step by only $1.8\times$. With the high-bandwidth MNNVL interconnect, all variants scaled well to 64 GPUs, and the greater arithmetic intensity of the GEMM-heavy configurations again led to higher parallel efficiency. At that scale, MNNVL was $1.8$--$2.6\times$ faster than the conventional inter-node configuration. Pressure accounted for $51$--$67\%$ of the Navier--Stokes runtime with MNNVL and $69$--$72\%$ without it; in weak scaling, its share reached $89$--$90\%$. The advantage of the faster interconnect is therefore not confined to the Poisson kernel, but carries over to the flow solver as a whole. These characteristics make incompressible-flow solvers well suited to rack-scale systems combining high-bandwidth communication with efficient dense linear algebra.

The additional cost of the dense transforms must ultimately be judged against the resolution saved by mesh stretching. The measured CPU timing ratios imply that, for the present problem sizes, if an FFT-based reference mesh uses uniform spacing comparable to the smallest required scale, reducing the total number of cells by roughly a factor of $2$--$3$ would make the fully GEMM-based case competitive in time-to-solution. The estimated break-even reduction on GPUs is of the same order for the Poisson solve and smaller for a complete Navier--Stokes step. A hybrid formulation offers the natural compromise: GEMM-based transforms can be reserved for directions in which stretching provides substantial resolution savings, while FFTs remain preferable where uniform spacing is adequate.

The solver can also be used in complex-flow methods with constant-coefficient elliptic operators. Examples include the immersed-boundary formulations in Refs.~\cite{Fadlun-et-al-JCP-2000,Breugem-and-Boersma-PoF-2005,Uhlmann-JCP-2005}. Related separable direct solvers have also been used in finite-difference schemes for cylindrical \cite{Verzicco-and-Orlandi-JCP-1996} and spherical \cite{Santelli-et-al-JCP-2021} coordinates, where Fourier diagonalization in a homogeneous angular direction is combined with a direct reduced-dimensional elliptic solve. Extending the present numerical-eigenbasis construction to more general curvilinear grids would require retaining a separable metric structure and modifying the corresponding one-dimensional operators.

Variable-coefficient Poisson problems, including those arising in multiphase and variable-property flows and in some embedded-boundary methods \cite{Gibou-et-al-JCP-2002,Mittal-et-al-JCP-2008}, may destroy separability and thereby break the required tensor-product structure. Some formulations can instead recast these problems in constant-coefficient form, in which case the present solver may remain applicable; see, for example, Refs.~\cite{Dong-and-Shen-JCP-2012,Demou-et-al-IJHMT-2019,Yildiran-et-al-JCP-2024}.

Other extensions concern time-dependent operators and extreme-scale implementation. Time-varying operators, for example those arising from unsteady Robin boundary conditions, would require efficient updating of the one-dimensional eigendecompositions. At very large transform dimensions, storing dense eigenvector matrices and applying them through local GEMMs may also become memory-limited. This bottleneck could be mitigated by distributing the matrices and projections using the Scalable Universal Matrix Multiplication Algorithm (\texttt{SUMMA}) \cite{deGeijn-et-al-Concurrency}, as available in libraries including \texttt{ScaLAPACK} and \texttt{cuBLASMp}.

Taken together, these results demonstrate that the direct tensor-product framework provides an efficient and scalable route to large-scale incompressible-flow simulations on Cartesian grids stretched in multiple directions, retaining FFTs where uniform spacing is adequate and using numerical eigenbasis transforms where stretching is required.

\section*{Acknowledgments}

Pedro Costa thanks Dr.~Mathieu Pourquie for insightful discussions on the symmetry of the finite-volume flavor of the three-dimensional Laplace operator. The CPU scaling tests on the supercomputer Snellius (based at SURF, The Netherlands) were supported by the NWO Grant no.\ \texttt{2024/ENW/01704792}, and SURF grant no.\ \texttt{EINF-17505}.

\bibliographystyle{elsarticle-num-nodoi}
\bibliography{bibfile}

\end{document}